\documentclass{eptcs}
\usepackage[utf8]{inputenc} 

\usepackage[american]{babel} 
\usepackage{csquotes} 

\usepackage{graphicx}
\usepackage{tikzscale}
\usepackage{caption}
\graphicspath{{./figures/}}

\usepackage{booktabs}

\usepackage{mathtools}
\usepackage{siunitx}
\usepackage{amssymb}

\usepackage{cite}

\usepackage[utf8]{inputenc}
\usepackage{subfiles}
\usepackage{comment}
\usepackage{mathtools}
\usepackage{inputenc}
\usepackage{subcaption}
\usepackage[usenames,dvipsnames]{xcolor}
\usepackage{booktabs}
\usepackage{array}
\usepackage{pbox}
\usepackage[mode=buildnew]{standalone}
\usepackage{tikz}
\usepackage{graphicx}
\usepackage{hyperref}
\hypersetup{colorlinks,linkcolor={blue!50!black},citecolor={red!40!black},urlcolor={blue!75!black},linktoc=page}
\usepackage{amsthm}
\usepackage{mathrsfs}
\usepackage{paralist}
\usepackage{bussproofs}

\EnableBpAbbreviations

\usepackage{bm}
\usepackage[capitalize,noabbrev]{cleveref}

\crefname{equation}{}{}
\crefname{condition}{condition}{conditions}
\crefname{item}{}{}
\crefname{section}{Chapter}{Chapters}
\crefname{subsection}{Section}{Sections}

\definecolor{mypurple}{rgb}{0.5, 0.0, 0.5}

\newcommand{\ca}{\mathbf}
\newcommand{\inp}{\mathrm{in}}
\newcommand{\out}{\mathrm{out}}

\newcommand{\RR}{\mathbb{R}}

\usetikzlibrary{positioning,decorations.pathreplacing,decorations.markings,arrows.meta,calc,fit,quotes,cd,math,arrows,backgrounds,shapes.geometric}

\theoremstyle{plain}

\theoremstyle{definition}

\theoremstyle{remark}

\newcommand{\ncat}[1]{\mathbf{#1}} 

\newcommand{\Set}{\ncat{Set}}
\newcommand{\Lin}{\ncat{Lin}}

\newcommand{\Cat}{\ncat{Cat}}

\newcommand{\SmallBox}[3]
{\begin{tikzpicture}[oriented WD, baseline=(Y.south), bb Small]
\node[inner sep=.1cm] [bb={1}{1}] (X) {$\scriptstyle #3$};
\draw[label] node[left=.1 of X_in1] (Y) {$#1$}
             node[right=.1 of X_out1] {$#2$};
\end{tikzpicture}}

\newcommand{\SmallBoxTwo}[4]
{\begin{tikzpicture}[oriented WD, baseline=(Y.south), bb Small]
\node[inner sep=.1cm] [bb={2}{1}] (X) {$\scriptstyle #4$};
\draw[label] node[left=.1 of X_in1] {$#1$}
node[left=.1 of X_in2] {$#2$}
node[right=.1 of X_out1] {$#3$};
\end{tikzpicture}}

\newcommand{\SmallBoxTwoOut}[4]
{\begin{tikzpicture}[oriented WD, baseline=(Y.south), bb Small]
\node[inner sep=.1cm] [bb={1}{2}] (X) {$\scriptstyle #4$};
\draw[label] node[left=.1 of X_in1] {$#1$}
node[right=.1 of X_out1] {$#2$}
node[right=.1 of X_out2] {$#3$};
\end{tikzpicture}}

\tikzcdset{
   arrow style=tikz,
   diagrams={>={Classical TikZ Rightarrow[angle=63:4pt, line width=.6pt]}},
   arrows={semithick}
}

\tikzset{
  tick/.style={postaction={
    decorate,
    decoration={markings, mark=at position 0.5 with {\draw[-] (0,.4ex) -- (0,-.4ex);}}}
  },
  tickx/.style={
    postaction={ decorate,
      decoration={markings,
        mark=at position 0.5 with {
          \fill circle [radius=.28ex];
        }
      }
    }
  }
}

\tikzset{
   dom/.style={append after command={coordinate[alias=dom#1]}},
   domA/.style={dom=A}, domB/.style={dom=B},
   domC/.style={dom=C}, domD/.style={dom=D},
   domE/.style={dom=E}, domF/.style={dom=F},
   cod/.style={append after command={coordinate[alias=cod#1]}},
   codA/.style={cod=A}, codB/.style={cod=B},
   codC/.style={cod=C}, codD/.style={cod=D},
   codE/.style={cod=E}, codF/.style={cod=F}
}

\tikzset{
   oriented WD/.style={
      every to/.style={out=0,in=180,draw},
      label/.style={
         font=\everymath\expandafter{\the\everymath\scriptstyle},
         inner sep=0pt,
         node distance=2pt and -2pt},
      semithick,
      node distance=1 and 1,
      decoration={markings, mark=at position .5 with {\arrow{stealth};}},
      ar/.style={postaction={decorate}},
      execute at begin picture={\tikzset{
         x=\bbx, y=\bby,
         every fit/.style={inner xsep=\bbx, inner ysep=\bby}}}
      },
   bbx/.store in=\bbx,
   bbx = 1.5cm,
   bby/.store in=\bby,
   bby = 1.75ex,
   bb port sep/.store in=\bbportsep,
   bb port sep=2,
   bb port length/.store in=\bbportlen,
   bb port length=4pt,
   bb min width/.store in=\bbminwidth,
   bb min width=1cm,
   bb rounded corners/.store in=\bbcorners,
   bb rounded corners=2pt,
   bb small/.style={bb port sep=1, bb port length=2.5pt, bbx=.4cm, bb min width=.4cm, bby=.7ex},
   bb Small/.style={bb port sep=1, bb port length=2.5pt, bbx=.5cm, bb min width=.5cm, bby=1ex},
   bb/.code 2 args={
      \pgfmathsetlengthmacro{\bbheight}{\bbportsep * (max(#1,#2)+1) * \bby}
      \pgfkeysalso{draw,minimum height=\bbheight,minimum width=\bbminwidth,outer sep=0pt,
         rounded corners=\bbcorners,thick,
         prefix after command={\pgfextra{\let\fixname\tikzlastnode}},
         append after command={\pgfextra{\draw
            \ifnum #1=0{} \else foreach \i in {1,...,#1} {
               ($(\fixname.north west)!{\i/(#1+1)}!(\fixname.south west)$) +(-\bbportlen,0) coordinate
               (\fixname_in\i) -- +(\bbportlen,0) coordinate (\fixname_in\i')}\fi 
            \ifnum #2=0{} \else foreach \i in {1,...,#2} {
               ($(\fixname.north east)!{\i/(#2+1)}!(\fixname.south east)$) +(-\bbportlen,0) coordinate
               (\fixname_out\i') -- +(\bbportlen,0) coordinate (\fixname_out\i)}\fi;
         }}}
   },
   bb name/.style={append after command={\pgfextra{\node[anchor=north] at (\fixname.north) {#1};}}}
}

\title{Compositional Cyber-Physical Systems Modeling}

\author{Georgios Bakirtzis
\institute{University of Virginia\\ Charlottesville, VA USA}
\email{bakirtzis@virginia.edu}
\and
Christina Vasilakopoulou
\institute{University of Patras\\ Patras, Greece}
\email{cvasilak@math.upatras.gr}
\and
Cody H. Fleming
\institute{University of Virginia\\ Charlottesville, VA USA}
\email{fleming@virginia.edu}
}
\def\titlerunning{Compositional Cyber-Physical Systems Modeling}
\def\authorrunning{G. Bakirtzis, C. Vasilakopoulou \& C.H. Fleming}

\begin{document}

\maketitle

\begin{abstract}
Assuring the correct behavior
of cyber-physical systems requires significant modeling effort, particularly during early stages
of the engineering and design process when a system is not yet available
for testing or verification of proper behavior.
A primary motivation for `getting things right'
in these early design stages is that altering the design is significantly less costly
and more effective than when hardware
and software have already been developed.
Engineering cyber-physical systems requires the construction
of several different types of models, each representing a different view,
which include stakeholder requirements, system behavior,
and the system architecture.
Furthermore, each of these models can be represented
at different levels of abstraction.
Formal reasoning has improved the precision
and expanded the available types of analysis
in assuring correctness of requirements, behaviors, and architectures.
However, each is usually modeled in distinct formalisms
and corresponding tools.
Currently, this disparity means that a system designer must manually check 
that the different models are in agreement. Manually editing 
and checking models is error prone, time consuming,
and sensitive to any changes in the design of the models themselves.
Wiring diagrams and related theory
provide a means
for formally organizing these different
but related modeling views, resulting
in a compositional modeling language
for cyber-physical systems.
Such a categorical language can make concrete the relationship between different model views,
thereby managing complexity, allowing hierarchical decomposition of system models,
and formally proving consistency between models.
\end{abstract}

\section{Introduction}

Cyber-physical systems are composed of computing platforms, control systems, sensors,
and communication networks~\cite{rajkumar:2010}. These systems provide critical service capabilities
in a number of engineering domains, including transportation, medical devices,
and power, to name a few.
The design of cyber-physical systems poses new challenges because
of the intertwined nature of digital control
with physical processes and the environment.
As autonomy and coordination between a multitude
of such systems becomes commonplace, there is an increasing need
to formally assure not only their individual behavior
but also the emergent properties that arise from the behavior of the composite system.
One such emergent property is safety. 
When cyber-physical systems exhibit unwanted behaviors, they can transition
to hazardous states and then lead to accidents.
To avoid such undesirable outcomes it is necessary
to provide evidence of correct behavior
before deployment, during the design phase of the systems lifecycle.
Design changes in later stages of the system lifecycle cost more
and are less effective~\cite{saravi:2008}.
It is only possible to produce this evidence early 
-- when an implemented system is not yet available -- 
through the management and use of various models.

The formal assurance of such complex systems requires the use
of compositional methods to manage the various views necessary
for the modeling and simulation of cyber-physical system design artifacts~\cite{duran:2020}.
At the highest level of abstraction a system must be constrained
by requirements, which define what the system ought to and ought not to do.
Through an iterative process, requirements are refined
to dictate the set of allowable behaviors 
and then used to architect a candidate implementation.
There are often several candidate implementations
that are functionally equivalent
but implemented with different hardware or software and they can be 
 compared in terms of how they respond
to different performance metrics.

There is a gap in systems engineering,
including its currently narrowly overlapping fields
of safety and security, where there is a lack of semantic strength\footnote{The complement of semantic strength is a perhaps familiar term: semantic gap.}
between requirements, behaviors,
and architectures.
These are considered connected in practice
but currently reside in distinct formalisms, leading to an unmanageable problem
of complexity~\cite{gell-mann:1994,gell-mann:2010}.
However, each is naturally modeled in different formalisms, 
for example,  first-order logic for requirements, difference and differential equations 
for physical behavior, and graphs for architectures.
Indeed, state space representations in control are informally related
to finite state machines in the software implementation.
Compositional formal methods assist with making this informal
understanding of relation and unification explicit.

Recent work in applied category theory for dynamical systems~\cite{fong:2018,spivak:2016,spivak:2020},
probabilistic failure analysis~\cite{breiner:2019}, control~\cite{baez:2015}, power systems~\cite{nolan:2019}, and requirements management~\cite{kibret:2019}
is particularly relevant to making this unification happen.
Specifically, the wiring diagram category~\cite{WiringDiagram} is useful 
for compositionally modeling cyber-physical systems
because it models reactive systems, feedback, and real-time computation,
all of which are necessary properties for modeling cyber-physical systems~\cite[\S 1.2]{Alur:2015}.
However, to date there is little work in merging this mathematical machinery
for the design of a modeling language blueprint for cyber-physical systems
using categorical, or otherwise algebraic, concepts.

Our aim with this line of research is
to use categorical formalisms and concepts
to better model systems for the purposes of system assurance.
To achieve this aim we bring theoretical concepts
from applied category theory to the realm of application
by modeling and simulating cyber-physical systems in a categorical context.
This requires us to precisely unify a number of diverse models
for the purpose of having a holistic picture of the systems behavior
and how it relates to candidate implementations.

We provide evidence
that using category theory as a modeling language
makes it possible to unify the necessary but diverse views
of cyber-physical system models. 
Using category theory solves an important problem in systems theory
\emph{and} systems engineering practice, which is a lack of formal traceability
between requirements, behaviors, and implementations.
We posit that wiring diagrams provide 
an effective solution to formal unification of system models 
that will improve scalability of the overall modeling effort.
As a first step, it is required that we reverse engineer 
the composite behavior of a system categorically,
and combine these possible decompositions with appropriate hierarchical implementations to provide flexibility
for the system designer: this is the main contribution of this paper.
The results of this ongoing project, which uses real-life systems and design needs, exhibit the usefulness of category theory
for finding consistency errors in system models early in the design phase. Applying formal tools and constructions directly to the field of systems engineering results 
in highly unified engineering design artifacts, which in turn have the potential to reduce the amount of faulty and dangerous systems that get deployed.

\section{Categorical Semantics for Cyber-Physical Systems}

In engineering there are two main iterative design steps, \emph{synthesis} and \emph{analysis}.
Synthesis is the process by which new systems are formed
from already known device laws,
and analysis is the process by which the new system is evaluated as a whole
for its overall function.
At each iteration of synthesis and analysis there is a notion
of \emph{refinement}.
Designing systems starts at some level
of abstraction and as we progress through the design-phase the modeling artifacts are developed
to be closer and closer to the eventual real system that will be deployed. 

In this work, we follow the categorical formalism of wiring diagrams~\cite{WiringDiagram,spivak:2016}. After we introduce the relevant theory in this section, we proceed to investigating a specific example of a cyber-physical system, namely an unmanned aerial vehicle (UAV). We start with modeling the system functions (Section~\ref{sec:simulation}) and iteratively decompose down to an implementation (Section~\ref{sec:implementation}) of an actual flight controller and airframe.
We show that by using wiring diagrams we can synthesize system models, analyze them with respect to their behavioral outcomes, and refine them from a particular set of behaviors to a concrete candidate architecture.

Recall that for a category $\ca{C}$ with products, the category $\ca{W}_\ca{C}$ of \emph{labeled boxes} 
and \emph{wiring diagrams}\footnote{The original definition involves $\ca{C}$-\emph{typed finite sets} and functions that respect the types; herein we abuse the notation by referring to the resulting category by taking products of types~\cite[Rem.~2.2.2]{spivak:2016}.} has as objects pairs of objects $X=(X_\inp,X_\out)$ that should be thought
of as the input and output types of an interface
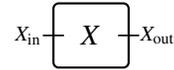
\begin{center}
\begin{tikzpicture}[oriented WD, bbx=.1cm, bby =.1cm, bb port sep=.15cm]
	\node [bb={1}{1}] (X) {$X$};
	\draw[label]
		node[left=.1 of X_in1]  {$X_\inp$}
        node[right=.1 of X_out1] {$X_\out$};
\end{tikzpicture}
\end{center}
that serves as a placeholder for some system,
whereas a morphism $X\to Y$ is a pair of arrows $( f_\inp\colon X_\out\times Y_\inp\to X_\inp, f_\out\colon X_\out\to Y_\out
)$ in $\ca{C}$ that should be thought of as providing the flow of information in a picture as follows
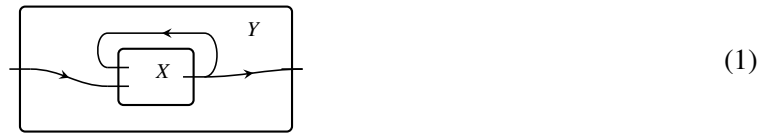
\begin{equation}\label{eq:wirdiagpic}
    \begin{tikzpicture}[oriented WD,baseline=(Y.center), bbx=2em, bby=1.2ex, bb port sep=1.2]
\node[bb={2}{1}] (X) {};
\node[bb={1}{1}, fit={($(X.north east)+(0.7,1.7)$) ($(X.south west)-(.7,.7)$)}] (Y) {};
\draw[ar] (Y_in1') to (X_in2);
\draw[ar] (X_out1) to (Y_out1);
\draw[ar] let \p1=(X.north west), \p2=(X.north east), \n1={\y1+\bby}, \n2=\bbportlen in
	(X_out1) to[in=0] (\x2+\n2,\n1) -- (\x1-\n2,\n1) to[out=180] (X_in1);
\draw [label] node at ($(Y.north east)-(.5cm,.3cm)$) {$Y$} node at ($(X.north east)-(.4cm,.3cm)$) {$X$}
node[above=of Y.north] {};
\end{tikzpicture}
\end{equation}
It is a monoidal category, with $X\otimes Y=(X_\inp\times Y_\inp, X_\out\times Y_\out)$  expressing parallel placement
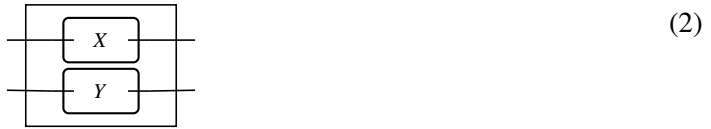
\begin{equation}\label{eq:tensorpic}
\begin{tikzpicture}[oriented WD,baseline=(Y.center), bbx=1.3em, bby=1ex, bb port sep=.06cm]
 \node[bb={1}{1}] (X1) {};
 \node[bb={1}{1},below =.5 of X1] (X2) {};
\node[fit=(X1)(X2),draw] {};
 \draw[label] 
              node at ($(X1.west)+(1,0)$) {$X$}
              node at ($(X2.west)+(1,0)$) {$Y$};
 \draw (X1_in1) -- (-2.5,0);
 \draw (X1_out1) -- (2.5,0);
  \draw (X2_in1) -- (-2.5,-3.85);
 \draw (X2_out1) -- (2.5,-3.85);
 \end{tikzpicture}
\end{equation}
In what follows, $\ca{C}$ will be $\Set$ or $\Lin$ (the category of linear spaces and linear maps) whereas in general the types could be objects in any cartesian monoidal category, which provides the flexibility to model different sorts of information or views of systems as well as incorporate notions of discrete and continuous time or functorially map between systems with different values. We sometimes write just $\ca{W}$ for $\ca{W}_\ca{C}$.

\begin{figure}[!ht]
\centering
\includegraphics[width=.7\linewidth]{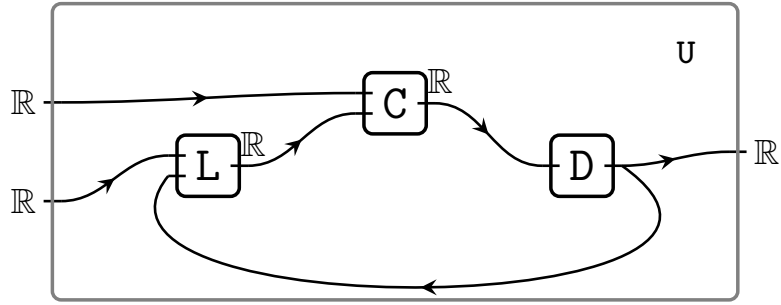}
\caption{Wiring diagrams make explicit the type of interconnections between subsystems.}\label{fig:UAVwiringdiagram}
\end{figure}

\begin{figure}[!ht]
\centering
\includegraphics[width=.85\linewidth]{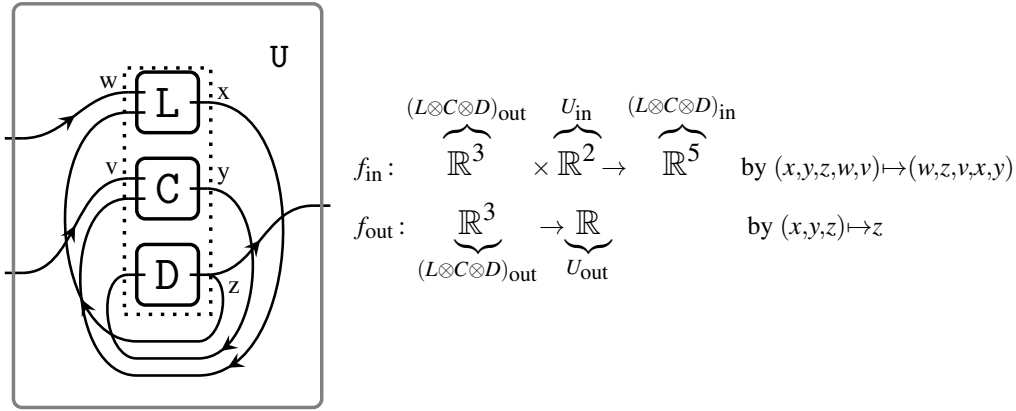}
\caption{In wiring diagrams the syntax defines the architecture; that is, how the boxes are arranged.}
\label{wirdiag}
\end{figure}

Morphisms in this category give a concrete mathematical description of pictures like \cref{fig:UAVwiringdiagram} where the interconnection of three systems $\mathtt{L},\mathtt{C},\mathtt{D}$ produces a composite system $\mathtt{U}$. As objects in $\ca{W}_\Set$, these are $\mathtt{L}=\mathtt{C}=\mathtt{U}=(\mathbb{R}^2,\mathbb{R})$ and $\mathtt{D}=(\mathbb{R},\mathbb{R})$, whereas the morphism $\mathtt{L}\otimes\mathtt{C}\otimes\mathtt{D}\to\mathtt{U}$ that captures the above arrangement is explicitly given in \cref{wirdiag}.

The essence of the proposed categorical formalism for systems modeling is encapsulated by viewing different types of system behaviors or requirements as \emph{algebras} for $\ca{W}$, namely lax monoidal functors\footnote{These are in fact usually pseudofunctors and more specifically \emph{monoidal indexed categories}~\cite[Rem.~2.1.1]{spivak:2016}.} from $\ca{W}$ to $(\Cat,\times,\mathbf{1})$. So each algebra assigns to a box $X=(X_\inp,X_\out)$ a category $FX$ of systems that can be placed inside the box, and also assigns to a wiring diagram $(f_\inp,f_\out)\colon X\to Y$ a functor $FX\to FY$ that, given a system inhabiting the internal box, produces the \emph{composite} system inhabiting the external box. 
Moreover, the laxator $\phi_{X,Y}\colon FX\times FY\to F(X\otimes Y)$ maps two given systems inside the boxes in wiring diagram \cref{eq:tensorpic} to a system inside their parallel placement.

One of the examples used in this work is the algebra of \emph{linear time-invariant systems}, also called \emph{linear dynamical systems} by Spivak~\cite{SpivakSteadyStates}. The algebra, namely a lax monoidal functor
\begin{equation}\label{eq:LTIS}
  \mathrm{LTIS}\colon\mathbf{W}_{\Lin}\to\Cat  
\end{equation}
assigns to any box $\SmallBox{X_\inp}{X_\out}{}$ a special discrete dynamical system; that is, a set $S$ called the \emph{state space} and two functions $u,r$ called \emph{update} and \emph{readout} as in
$$(S,u\colon S\times X_\inp\to S,r\colon S\to X_\out)$$ 
where all $S,X_\inp,X_\out$ are linear spaces and both update and readout functions $u$ and $r$ are linear functions expressed as
\begin{displaymath}
u(s,x)= \mathscr{A}\cdot s+ \mathscr{B}\cdot x=\begin{pmatrix}\mathscr{A} & \mathscr{B}
\end{pmatrix}
\begin{pmatrix}
s \\
x
\end{pmatrix}\qquad
r(s)= \mathscr{C}\cdot s
\end{displaymath}
where $\mathscr{A}$, $\mathscr{B}$ and $\mathscr{C}$ are matrices of appropriate dimension and $\cdot$ is matrix multiplication. For example, if $X_\inp=\RR^k$, $X_\out=\RR^\ell$ and $S=\RR^n$, then $\mathscr{A}\in {}_nM_n$ represents a transformation $\RR^n\to\RR^n$, $\mathscr{B}\in {}_nM_k$ a transformation $\RR^k\to\RR^n$ and $\mathscr{C}\in {}_\ell M_n$ a transformation $\RR^n\to\RR^\ell$. We often write $\stackrel{\bullet}{s}=u(s,x)$ as a standard notation, even in this discrete time-case. 

Now given an arbitrary wiring diagram $(f_\inp, f_\out)\colon (X_\inp,X_\out)\to (Y_\inp,Y_\out)$, notice that for $Y_\inp=\RR^{k'}$ and $Y_\out=\RR^{\ell'}$ both components of the wiring diagram are also expressed via matrices $$F_\inp=\begin{pmatrix}
{}_k(\mathscr{A}^F)_{\ell} & {}_k(\mathscr{B}^F)_{k'}
\end{pmatrix} \qquad
F_\out= {}_{\ell'}\mathscr{C}^F_{\ell}$$ The functor $\textrm{LTIS}$ maps some system $(S,\mathscr{A},\mathscr{B},\mathscr{C})$ in $\SmallBox{\RR^k}{\RR^\ell}{X}$ to the composite system 
\begin{equation}\label{eq:LTIScomposite}
(S,\mathscr{A}+\mathscr{B}\cdot \mathscr{A}^F\cdot \mathscr{C},\; \mathscr{B}\cdot \mathscr{B}^F,\; \mathscr{C}^F\cdot \mathscr{C})
\end{equation}
in $\SmallBox{\RR^{k'}}{\RR^{\ell'}}{Y}$. The laxator of $\mathrm{LTIS}$ maps any two such systems $(S_1,\mathscr{A}_1,\mathscr{B}_1,\mathscr{C}_1)$,  $(S_2,\mathscr{A}_2,\mathscr{B}_2,\mathscr{C}_2)$ inhabiting parallel boxes to the system 
\begin{displaymath}
\left(S_1\times S_2,\begin{pmatrix}
\mathscr{A}_1 & 0 \\
0 & \mathscr{A}_2
\end{pmatrix},
\begin{pmatrix}
\mathscr{B}_1 & 0 \\
0 & \mathscr{B}_2
\end{pmatrix},
\begin{pmatrix}
\mathscr{C}_1 & 0 \\
0 & \mathscr{C}_2
\end{pmatrix}\right).
\end{displaymath}

This is the machinery that will allow us to model control systems
and, therefore, the largest class of cyber-physical systems.

\section{Simulating Dynamics}\label{sec:simulation}

A UAV is often a small factor fixed wing plane or multi-rotor copter
that is controlled on the ground by a computer.
The unmanned aerial system, which includes the UAV, ground control station,
and a network between the two, is responsible
for navigating the plane from one waypoint to the next.

A UAV includes a control system and as such it needs to adhere
to certain control laws, which have to be designed, implemented,
and tuned.
But what is a \emph{behavioral} interpretation
of what a UAV does?
We can consider behavior as a composition of system functions.
In this sense, a compositional interpretation could be
the composition of its sensors, controllers, and control surfaces
to produce a known set of dynamics.
To simulate the UAV behavior the first step is to produce the dynamic response
of the system to environmental values.
We capture the dynamics 
of such a system to the systems-as-algebras paradigm,
where we consider that the sensors are measuring the environment
but also get input from the resulting airframe dynamics. We describe this process step-by-step in a practical approach, and conclude the section with the bare categorical dimension of said process.

\begin{figure}[!t]
\centering
\includegraphics[width=.9\linewidth]{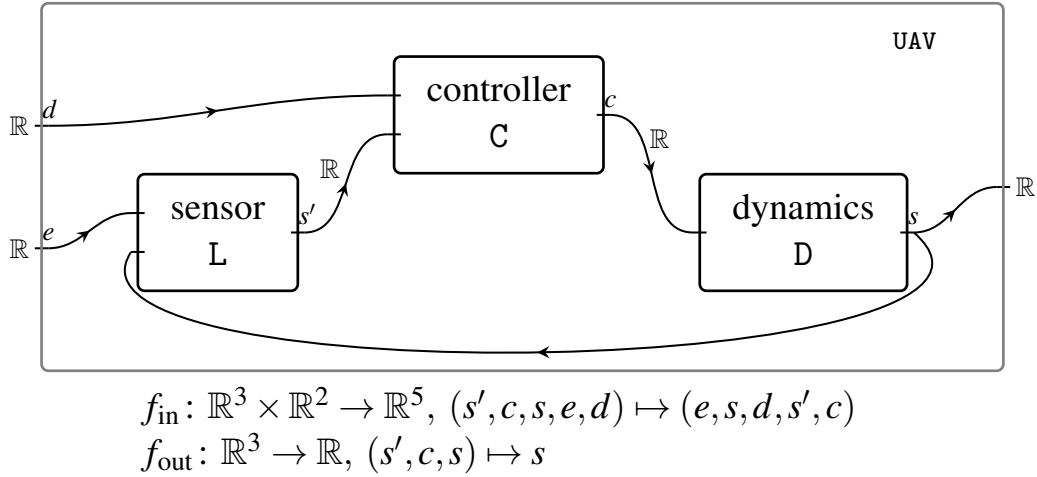}
\caption{The physical decomposition of the unmanned aerial vehicle (UAV), where $d$ denotes the desired state, $s'$ the predicted state, $c$ the control action, $s$ the current state, and $e$ the environmental inputs. The functions coincide with $\mathtt{U}$ (\cref{wirdiag}).}\label{fig:system_function}
\end{figure}

Thinking of the composite UAV system in terms of its control laws for discrete time, we would model its dynamics by its difference equations -- since we are operating in discrete time -- but we map the matrix representation as above. Therefore
\begin{equation*}
    s_{p+1} = \mathscr{A}s_p + \mathscr{B} c_p
\end{equation*}
\noindent
where $s_p \in \mathbb{R}^n$ is the discrete time state and 
$c_p \in \mathbb{R}^n$ is the control signal/output, and
\begin{equation*}
y_p = \mathscr{C}s_p + \mathscr{D}c_p
\end{equation*}
is the measurement which is also $y_{p} \in \mathbb{R}^n$, where we could assume a perfect measurement so that $s_p = y_p$. In practice, the airframe behavior (say, the static parts of the airframe) can be considered or otherwise contained in $\mathscr{A}$, servos and actuators would be part
of $\mathscr{B}$, sensors are part of $\mathscr{C}$. 
The disturbance, or feedforward term, $\mathscr{D}$ is typically $0$ of proper dimension.
We define the following notational abbreviations:
\begin{align*}
\mathtt{L} & \triangleq \text{ sensor,}\\
\mathtt{C} & \triangleq \text{ controller, and}\\
\mathtt{D} & \triangleq \text{ dynamics.}
\end{align*}
Any such system requires a prediction $s'$, this is typically used
within $\mathtt{C}$ (Figure~\ref{fig:system_function}).
This is done in different ways (open-loop, MPC, classical control, et cetera).
In modern control, one typically takes a measurement $y_p$ and uses it to create an estimate of the full state at that time $s_{p+1}$, 
and then uses that state information to update its control signal. 

At this point, we have mapped a given behavioral formalism
to our wiring diagram syntax and semantics for illustrative purposes.
We posit that such diagrams can fully capture behaviors that adhere to composition.
A different formalism could model the behavior
of an information technology (IT) system
by mapping a graph formalism used to assess and measure its performance.
However, the usefulness of the formalism lies precisely
in relating the physical behavior of the cyber-physical systems
to its candidate architecture and therefore composition
in this sense better captures the possible misbehavior
a system designer might want to mitigate against.

We now wish to use the compositional system-as-algebra perspective to, in a sense, reverse engineer: given a specific behavioral description of some UAV we would like to provide specific descriptions for the individual component systems that would result in such a composite behavior. We will use the system model for predicting aircraft pitch~\cite{systemModel}. In that specific example, the longitudinal equations of motion for a fixed-winged aircraft are given as a state-space model of the following form,
\begin{equation}\label{eq:boeing}
 \begin{pmatrix}
  \stackrel{\bullet}{a} \\
  \stackrel{\bullet}{q} \\
  \stackrel{\bullet}{\theta}
 \end{pmatrix}=
 \begin{pmatrix}
-0.313 & 56.7 & 0 \\
-0.0139 & -0.426 & 0 \\
0 & 56.7 & 0
 \end{pmatrix}
 \begin{pmatrix}
a \\
q \\
\theta
 \end{pmatrix}+
 \begin{pmatrix}
0.232 \\
0.0203 \\
0
 \end{pmatrix}\begin{pmatrix}\delta\end{pmatrix}
\end{equation}
\begin{displaymath}
 y=\begin{pmatrix}0 & 0 & 1\end{pmatrix}\begin{pmatrix}a \\ q \\ \theta \end{pmatrix}
\end{displaymath}
where $a$ is the angle of attack, $q$ is the pitch rate, $\theta$ is the pitch angle and $\delta$ is the deflection angle.

Working in the linear time-invariant system algebra \cref{eq:LTIS}, suppose \((S_\mathtt{L},\mathscr{A}_\mathtt{L},\mathscr{B}_\mathtt{L},\mathscr{C}_\mathtt{L})\), \((S_\mathtt{C},\mathscr{A}_\mathtt{C},\mathscr{B}_\mathtt{C},\mathscr{C}_\mathtt{C})\) and \((S_\mathtt{D},\mathscr{A}_\mathtt{D},\mathscr{B}_\mathtt{D},\mathscr{C}_\mathtt{D})\) are three linear systems inhabiting the respective boxes of \cref{fig:system_function}, with
\begin{align*}
u_\mathtt{L}(s_\mathtt{L},e,s)=&\mathscr{A}_\mathtt{L}\cdot s_L+ \mathscr{B}_L\cdot(e,s) & 
r_\mathtt{L}(s_\mathtt{L})=&\mathscr{C}_\mathtt{L}\cdot s_\mathtt{L} \\
u_\mathtt{C}(s_\mathtt{C},d,s')=&\mathscr{A}_\mathtt{C}\cdot s_\mathtt{C}+ \mathscr{B}_\mathtt{C}\cdot(d,s') &
r_\mathtt{C}(s_\mathtt{C})=&\mathscr{C}_\mathtt{C}\cdot s_\mathtt{C} \\
u_\mathtt{D}(s_\mathtt{D},c)=&\mathscr{A}_\mathtt{D}\cdot s_\mathtt{D}+ \mathscr{B}_\mathtt{D}\cdot(c) &
r_\mathtt{D}(s_\mathtt{D})=&\mathscr{C}_\mathtt{D}\cdot s_\mathtt{D}
\end{align*}
Using the algebra machinery \cref{eq:LTIScomposite}, we can compute the composite linear dynamical system that inhabits the box $\mathtt{UAV}$: its state space is $S_\mathtt{L}\times S_\mathtt{C}\times S_\mathtt{D}$, and its update and readout linear functions are
\begin{align*}
u_{\mathtt{UAV}}\colon  S_\mathtt{L}\times S_\mathtt{C}\times S_\mathtt{D}\times\RR^2&\to S_\mathtt{L}\times S_\mathtt{C}\times S_\mathtt{D}, \\
 (s_\mathtt{L},s_\mathtt{C},s_\mathtt{D},d,e)&\mapsto(\mathscr{A}_\mathtt{L}s_\mathtt{L}+\mathscr{B}_\mathtt{L}\begin{pmatrix}e \\ \mathscr{C}_\mathtt{D}s_\mathtt{D}\end{pmatrix}, \mathscr{A}_\mathtt{C}s_\mathtt{C}+\mathscr{B}_\mathtt{C}\begin{pmatrix}d \\ \mathscr{C}_\mathtt{L}s_\mathtt{L}\end{pmatrix},\mathscr{A}_\mathtt{D}s_\mathtt{D}+\mathscr{B}_D\mathscr{C}_\mathtt{C}s_\mathtt{C}) \\
r_{\mathtt{UAV}}\colon  S_\mathtt{L}\times S_\mathtt{C}\times S_\mathtt{D}&\to \RR \\
 (s_\mathtt{L},s_\mathtt{C},s_\mathtt{D})&\mapsto\mathscr{C}_\mathtt{D}s_\mathtt{D}
\end{align*}
We assume, for simplicity\footnote{In the likely case that sensor and controller are linear functions, to express them as linear time-invariant systems the state space would match the input space and therefore be $\mathbb{R}^2$. Moreover $\mathscr{A}$ would be the zero matrix and $\mathscr{B}$ would be the unit matrix.}, that the state spaces of the sensor and controller are $\mathbb{R}$. Knowing that only the dynamics $\mathtt{D}$ relate to the triplet $(a,q,\theta)$, we set $S_\mathtt{D}=\mathbb{R}^3$ producing a resulting state space of the composite system $S_\mathtt{UAV}=\mathbb{R}^5$. 
Moreover, from the shape of the boxes we deduce that matrices $\mathscr{A}_\mathtt{L}$, $\mathscr{A}_\mathtt{C}$, $\mathscr{C}_\mathtt{L}$ and $\mathscr{C}_\mathtt{C}$ are one-by-one, $\mathscr{B}_\mathtt{L}$ and $\mathscr{B}_\mathtt{C}$ are one-by-two, whereas $\mathscr{A}_\mathtt{D}$ is three-by-three, $\mathscr{B}_\mathtt{D}$ is three-by-one and $\mathscr{C}_\mathtt{D}$ is one-by-three. 

Decoding the above equations, we first of all have that the only output of the composite system is the output of $\mathtt{D}$, namely
\begin{displaymath}
\mathscr{C}_\mathtt{UAV}=\begin{pmatrix}0 & 0 & _1(\mathscr{C}_\mathtt{D})_3
\end{pmatrix}
\end{displaymath}
Hence for obtaining equation \cref{eq:boeing}, in the specific example we deduce that $\mathscr{C}_\mathtt{D}=\begin{pmatrix}0 & 0 & 1
\end{pmatrix}$ meaning only $\theta$ is outputted to the outside world as desired.

Regarding the update part for an element of the state space $\RR^5$ of the form $(s_\mathtt{L},s_\mathtt{C},\overbrace{a,q,\theta}^{s_\mathtt{D}})$, isolating the first two variables we obtain
\begin{displaymath}
\stackrel{\bullet}{s_\mathtt{L}}=\mathscr{A}_\mathtt{L}s_\mathtt{L}+{}_1(\mathscr{B}_\mathtt{L})_2\begin{pmatrix}e \\ \underbrace{\mathscr{C}_\mathtt{D}s_\mathtt{D}}_\theta\end{pmatrix}\qquad
\stackrel{\bullet}{s_\mathtt{C}}=\mathscr{A}_\mathtt{C}s_\mathtt{C}+{}_1(\mathscr{B}_\mathtt{C})_2\begin{pmatrix}d \\ \mathscr{C}_\mathtt{L}s_\mathtt{L}\end{pmatrix}
\end{displaymath}
which could be viewed as the `extra info' of the composite system relating to the behaviors of the sensor and controller, not appearing in equation \cref{eq:boeing} but part of the total system's behavior. Now isolating the last three variables we obtain a description
\begin{displaymath}
\begin{pmatrix}\stackrel{\bullet}{\alpha} \\
\stackrel{\bullet}{q} \\
\stackrel{\bullet}{\theta}
\end{pmatrix}=
{}_3(\mathscr{A}_\mathtt{D})_3\begin{pmatrix}\alpha \\
q \\
\theta
\end{pmatrix}+{}_3(\mathscr{B}_\mathtt{D})_1\mathscr{C}_\mathtt{C}s_\mathtt{C}
\end{displaymath}
Comparing with the desired equation \cref{eq:boeing}, the deflection angle $\delta$ is the output of the controller $\mathscr{C}_\mathtt{C}s_\mathtt{C}$ which  matches the physical reality, and the $\mathscr{A}_\mathtt{D}$, $\mathscr{B}_\mathtt{D}$ are completely determined by the composite description, namely
\begin{displaymath}
\mathscr{A}_\mathtt{D}= \begin{pmatrix}
-0.313 & 56.7 & 0 \\
-0.0139 & -0.426 & 0 \\
0 & 56.7 & 0
 \end{pmatrix}
\qquad
\mathscr{B}_\mathtt{D}= \begin{pmatrix}
0.232 \\
0.0203 \\
0
 \end{pmatrix}
\end{displaymath}
The remaining data $\mathscr{A}_{\mathtt{L},\mathtt{C}},\mathscr{B}_{\mathtt{L},\mathtt{C}},\mathscr{C}_{\mathtt{L},\mathtt{C}}$ depend on engineering and physical parameters. Imposing extra, realistic conditions on what these could be is part of ongoing work.

In conclusion, this reverse engineering process is categorically described as follows. Given the algebra $\mathrm{LTIS}$, a wiring diagram $f\colon\mathtt{L}\otimes\mathtt{C}\otimes\mathtt{D}\to\mathtt{U}$ in $\ca{W}_\Lin$ (\cref{fig:system_function}) as well as an object of the target category $\mathrm{LTIS}(\mathtt{U})$, namely a specific linear system as in equation \cref{eq:boeing} inhabiting the outside box $\mathtt{UAV}$, the goal is to find an object in the pre-image of the given system under the composite functor
\begin{equation}\label{eq:compositionfunction}
 \mathrm{LTIS}(\mathtt{L})\times\mathrm{LTIS}(\mathtt{C})\times\mathrm{LTIS}(\mathtt{D})\xrightarrow{\phi_{\mathtt{L},\mathtt{C},\mathtt{D}}}
 \mathrm{LTIS}(\mathtt{L}\otimes\mathtt{C}\otimes\mathtt{D})\xrightarrow{\mathrm{LTIS}(f)}\mathrm{LTIS}(\mathtt{U})\text{.}
\end{equation}
Clearly it is not expected that such a problem has a unique solution, but for example, in this specific case the system 
\begin{equation}\label{eq:Dsystem}
(S_\mathtt{D},\mathscr{A}_\mathtt{D},\mathscr{B}_\mathtt{D},\mathscr{C}_\mathtt{D})    
\end{equation}
 was completely determined by the composite system. Further work would aim to shed light on possible shapes of wiring diagrams that have better-behaved solutions under algebras of interest.

\section{Decomposing Hierarchically to an Implementation}\label{sec:implementation}

We have shown how to model a behavioral response of the system as it pertains to its physics.
But how might we connect this behavioral understanding
to a concrete implementation, in particular how can we zoom-in on each of the boxes to create an \emph{architectural implementation} that will yield the wanted behaviors?
Of course, there is no one solution for how to implement a cyber-physical system, so how might we categorically capture different architectural solution with an explicit relationship to the wanted set of behaviors?

Starting with a cyber-physical system from an engineer or designer point of view we now might want to model a candidate implementation. That means decomposing it into certain sub-components and using a specific wiring between them, following some choices based on the physical reality, experience, purpose and access to particular components at the time.
Having formalized arbitrary zoomed-in pictures of a system or even more abstractly an agnostic process interface where various descriptions could live on using the category of labeled boxes and wiring diagrams, gives us the chance to view the above process using the basic notion of a \emph{slice category} under an appropriate perspective.

Broadly speaking, for any category $\ca{C}$ and a fixed object $c\in\ca{C}$, we here think of $\ca{C}/c$ as containing all possible `design choices' for $c$ available to a system engineer; this formally models the possibility of implementing a system $c$ in multitudes of ways.
For example, in our setting a designer might choose to implement the flight control system -- a subsystem of the UAV -- either using a microcontroller or a field programmable gate array (FPGA). Similarly, the same applies to the rest of the system: there are a number of options for implementing the sensory system or dynamics.

In fact, we have already used a chosen implementation in the previous section. In more detail,
suppose we have a process inhabiting some box $\SmallBoxTwo{\mathbb{R}}{\mathbb{R}}{\mathbb{R}}{\mathtt{U}}$, which is an object of $\ca{W}$.
How can we decompose it into sub-processes, and how should they be interconnected to form the given system? All the possible decompositions can be thought of as the objects of the slice category $\ca{W}/{\SmallBoxTwo{\mathbb{R}}{\mathbb{R}}{\mathbb{R}}{\mathtt{U}}}$, therefore the architectural implementation is manifested by choosing one such object: for example, $f\colon\mathtt{L}\otimes \mathtt{C}\otimes \mathtt{D}\to \mathtt{U}$ is one such choice (\cref{fig:UAVwiringdiagram}), namely a wiring diagram with target $\mathtt{U}$.
Considering all possible subcomponents and wirings of a box could be thought of as the object-function of the functor $\ca{W}/(\textrm{-})\colon\ca{W}\to\Cat$.

To control a UAV we could implement a number of design solutions, each with different components
and setup.
For illustrative purposes we show one such candidate implementation, one
that models a real UAV system.
We do discuss possible deviations from this one design
and how they are modeled in the slice category.
For the design of a candidate implementation we use the following notational abbreviations:
\begin{align*}
\mathtt{I} & \triangleq \text{ inertial measurement unit,}\\
\mathtt{P} & \triangleq \text{ processor},\\
\mathtt{V} & \triangleq \text{ servos},\\
\mathtt{X} & \triangleq \text{ aileron,}\\
\mathtt{Y} & \triangleq \text{ rudder,}\\
\mathtt{Z} & \triangleq \text{ throttle,}\\
\mathtt{W} & \triangleq \text{ elevator, and}\\
\mathtt{F} & \triangleq \text{ airframe.}
\end{align*}

For example, a possible set of components for the implementation of the sensory system for a flight control system includes a gyroscope,
an accelerometer, and a pressure sensor, which are usually realized through an inertial measurement unit (IMU).
Because an IMU has accumulative error it is important
to either add redundancy (that is, use two IMU devices) or add a separate device (for example, a global positioning unit (GPS) to ping and correct this error in IMU measurements).
Current UAV systems include further functionality,
for example, by using differential pressure sensors
or adding a magnetometer to achieve further robustness 
during guided flight.
Here we will focus on one possible design of the UAV,
including implementation for the dynamics through control surfaces; that is, aileron, rudder, throttle, and elevator,
and airframe.
To manipulate the control surfaces we include servos, which are small motors attached to the control surfaces that move them based on the control laws that are implemented in the controller.

\begin{figure}[!ht]
\centering
\includegraphics[width=1\linewidth]{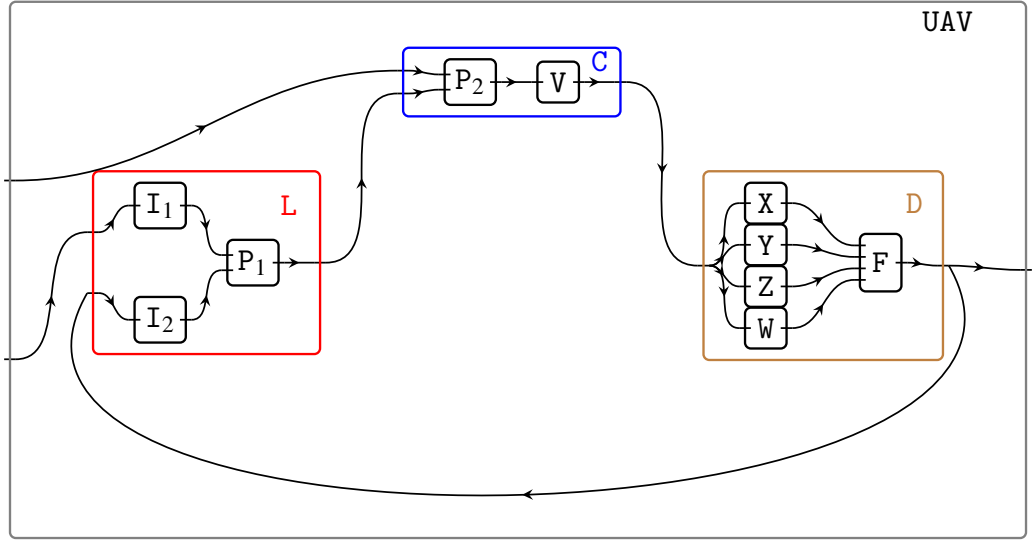}
\caption{The component systems are further decomposed to candidate implementations.}\label{fig:implementation}
\end{figure}

One of the important advantages of expressing system decompositions as a morphism in the category $\ca{W}$ is that we can perform further zoomed-in decompositions as desired, in a \emph{hierarchical} way. For example, we may choose to implement the sensor box $\mathtt{L}$ using two IMU units $\mathtt{I}_1,\mathtt{I}_2$ and a processor $\mathtt{P}_1$ in a certain interconnection. Expressing this as a morphism with target $\mathtt{L}$, namely $g\colon\mathtt{I}_1\otimes \mathtt{I}_2\otimes \mathtt{P}\to \mathtt{L}$ means that we can compose this with the earlier $f$ to obtain a two-level zoomed-in decomposition
\begin{displaymath}
(\mathtt{I}_1\otimes\mathtt{I}_2\otimes\mathtt{P}_1)\otimes \mathtt{C}\otimes \mathtt{D}\xrightarrow{g\otimes1\otimes1}\mathtt{L}\otimes \mathtt{C}\otimes \mathtt{D}\xrightarrow{f}\mathtt{UAV}.
\end{displaymath}
that only `opens-up' the box $\mathtt{L}$. We could moreover implement the control as well as the dynamics box, and decompose them in a choice of subcomponents and wires between them. An example where the control box is decomposed into $\mathtt{P}_2$ followed by $\mathtt{V}$ in a serial composition, and the dynamics box is decomposed into four parallel boxes, $\mathtt{X}$, $\mathtt{Y}$, $\mathtt{Z}$ and $\mathtt{W}$ followed by $\mathtt{F}$ amounts to choosing $h\colon \mathtt{P}_2\otimes\mathtt{V}\to \mathtt{C}$ in $\ca{W}/\SmallBoxTwo{}{}{}{\mathtt{C}}$ and $k\colon \mathtt{X}\otimes\mathtt{Y}\otimes\mathtt{Z}\otimes\mathtt{W}\otimes\mathtt{F}\to\mathtt{D}$ in $\ca{W}/\SmallBoxTwoOut{}{}{}{\mathtt{D}}$. Combining all these morphisms we have the composition (\cref{fig:implementation}):
\begin{displaymath}
(\mathtt{I}_1\otimes \mathtt{I}_2\otimes \mathtt{P}_1)\otimes (\mathtt{P}_2\otimes \mathtt{V})\otimes (\mathtt{X}\otimes \mathtt{Y}\otimes \mathtt{Z}\otimes \mathtt{W}\otimes \mathtt{F})\xrightarrow{g\otimes h\otimes k}\mathtt{L}\otimes \mathtt{C}\otimes \mathtt{D}\xrightarrow{f}\mathtt{UAV}
\end{displaymath}
that can be considered as a single morphism from the tensor of all second-level sub-components to $\mathtt{UAV}$, namely erasing the intermediate colored boxes.

A categorical approach to systems engineering allows us to hierarchically decompose
from the system function to a candidate implementation while making explicit the traceability between those two related by different views of system design.
Broadly speaking, this section's narrative relates to the architectural implementation of our systems, formally expressed by choosing the constituent boxes (domain) and their wiring interconnection (morphism) of a given system inside the category $\ca{W}$. The implementation comes together with the physical behavior of the UAV (Section \ref{sec:simulation}) via the algebra structure of any lax monoidal functor $B\colon\ca{W}\to\Cat$, in this instance $\mathrm{LTIS}$ \cref{eq:LTIS}, as follows.
For \emph{any}  $\mathtt{UAV}$-implementation $f\colon \SmallBox{}{}{A}\to\SmallBoxTwo{}{}{}{\mathtt{UAV}}$, the functorial assignment
\begin{displaymath}
\begin{tikzcd}[row sep=.1in]
B\colon\ca{W} \ar[r] & \Cat \\
\phantom{AB}\SmallBox{}{}{A}\ar[dd,shift left=2,"f"']\ar[r,mapsto] & B(\SmallBox{}{}{A})\ar[dd,"B(f)"] \\
&& \\
\phantom{AB}\SmallBoxTwo{}{}{}{\mathtt{UAV}}\ar[r,mapsto] & B(\SmallBoxTwo{}{}{}{\mathtt{UAV}})
\end{tikzcd}
\end{displaymath}
allows us to compose a given $B$-system on $\SmallBox{}{}{A}$ to produce a $B$-system on $\SmallBoxTwo{}{}{}{\mathtt{UAV}}$, or more interestingly go backwards: given a $B$-system on $\mathtt{UAV}$, we first choose an implementation $f\colon A\to\mathtt{UAV}$ where the domain could be the tensor of a number of sub-boxes, and then we choose an object of the system's pre-image similarly to the composite functor \cref{eq:compositionfunction}. This is a two-step process which appears in practice and depends on the choices of the system engineer, but can now be understood categorically through functorial semantics. 

In particular, earlier the linear system \cref{eq:Dsystem} inhabiting the box $\SmallBoxTwoOut{}{}{}{\mathtt{D}}$ was completely determined by the equations of the composite system, whereas now we could further elaborate on the possible linear systems inhabiting the boxes $\mathtt{X}$, $\mathtt{Y}$, $\mathtt{Z}$, $\mathtt{W}$ and $\mathtt{F}$. Moreover, if such possibilities do not agree with physical intuition or the expected service of the system, such an implementation may be ruled out and alternatives identified appropriately based on the allowable set of behaviors.

\section{Related Work}

Systems theory has as long tradition and spans a large number
of subfields, some more formal than others.
A formal framework for systems theory was first built
to describe biological systems~\cite{von:1950,ashby:1991}.
Since then, there have been several developments
to systems theory that have assisted 
in areas such as control~\cite{rasmussen:1985},
safety assessment~\cite{leveson:2004,leveson:2017},
and security analysis~\cite{young:2014,carter:2018}.
While we have been heavily influenced
by systems theory,
in this paper we develop a categorical approach 
to synthesizing and analyzing cyber-physical system designs
on top of systems theory.
This is because category theory is flexible
with respect to relating different abstraction levels,
which are necessary for the synthesis and analysis of cyber-physical systems.

The idea of using category theory
for unifying specification languages is not new~\cite{fiadeiro:1995,goguen:1991}.
However, we note that previous work in this area does not provide a concrete example
of the specified system and it has also been applied predominantly
for software engineering.
Cyber-physical systems require modeling both the dynamics
of the system and their implementation in software and hardware,
which is an approach we illustrate in this paper.
Further, such approaches to systems theory are old enough
to require presenting again with the developments
both in the theory and engineering of systems.
While the categorical language can potentially be used
for any system, we focus on its effectiveness to unify and scale models
of cyber-physical systems, which pose new challenges because they operate
in the environment.

In engineering too there has been research
that takes advantage of category theory.
A categorical formulation of hybrid systems
that unifies all its views was presented by Ames
and Sastry~\cite{ames:2005}.
Our work views dynamical systems at a different abstraction
than Ames and Sastry.
We formalize the different views of a cyber-physical system
as it would be modeled by systems engineers
instead of focusing to the different topologies studied in control,
which is often the case when designing systems in practice.
Furthermore, Hasuo presented the idea of a research program
for unifying the different views
of cyber-physical systems using category theory~\cite{hasuo:2017}.
Our work presents one (partial) solution to the problem outlined by Hasuo.

More recently, there have been other directions in the area
of a categorical language for system design and assessment~\cite{breiner:2019,breiner:2019a}, control~\cite{baez:2015},
and even requirements management~\cite{gebreyohannes:2018}.
We view these research directions as complementary
to the one presented in this paper.
We present a framework for cyber-physical systems,
which requires relating system assessment, modeling control,
and refining requirements.

\section{Conclusion}

Applied category theory is a new field
with promising solutions
to the formal unification and dissemination 
of engineering design artifacts.
This unification is of particular importance
for designing and assuring correctness of cyber-physical systems,
where misbehavior can lead to hazards.
Cyber-physical system models require a number of views
to provide evidence of correctness in their eventual deployment.
We have shown that it is possible
to create functorial relationships between the two main views
of a cyber-physical system, the simulation of the desired behavior
of the system and its decomposition to a specific candidate implementation.
By building this functorial relationship it is possible
to find consistency errors between those different model views
and, therefore, better address misbehavior early in the system's lifecycle.

This is a first step to an eventual unification not only
of methods in cyber-physical systems theory but also of tools.
This would be an important accomplishment given
that today systems are designed predominantly through tools.
For example, tools are used
to tune the control laws, simulate the overall system state, 
provide guarantees that the system does not reach death states,
and compare and contrast different candidate implementations.
We posit that working compositionally by formalizing cyber-physical systems modeling
in category theory will in the future lead to scalable models
of those complex systems.
In addition, applied category theory has already shown positive results
in the area of software execution, for example by using dependent types.
Producing a cyber-physical systems modeling language blueprint
in category theory could lead to having a fully integrated model
that could be used throughout the lifecycle.
In the beginning this model could be used to simulate the dynamical response
and towards the end it could be used to synthesize code from the model, which would bridge the gap between system specification and implementation.

Beyond our general goals for this research program, our main immediate improvements
to this work are to integrate discrete and continuous time,
extend notions from contract-based design~\cite{sangiovanni:2012} in category-theoretic terms,
and finally construct a fully traceable compositional model between requirements, behaviors,
and implementations of cyber-physical systems.


\section{Acknowledgments}

G. Bakirtzis and C.H. Fleming are partially supported through the Systems Engineering Research Center (SERC) under USDOD Contract HQ0034-13-D-0004, and through the Northrop Grumman Mission Systems’ University Research Program. Any opinions, findings and conclusions or recommendations expressed in this material are those of the authors and do not necessarily reflect the views of the USDOD nor Northrop Grumman. 

C. Vasilakopoulou would like to thank the General Secretariat for Research and Technology (GSRT) and the Hellenic Foundation for Research and Innovation (HFRI).

\bibliography{manuscript}
\bibliographystyle{eptcs}
\end{document}